\documentclass[12pt]{article}

\pdfoutput=1
\usepackage{scicite}
\usepackage{graphicx}
\usepackage{enumitem}
\usepackage{datetime}
\usepackage{gensymb}
\usepackage{scicite}
\usepackage{amsmath}
\usepackage{times}

\newenvironment{sciabstract}{%
\begin{quote} \bf}
{\end{quote}}

\title{Disorder Enhances the Fracture Toughness of Mechanical Metamaterials}

\author{
Sage Fulco,$^{1,\ast}$ Michal K. Budzik$^{2}$, Hongyi Xiao,$^{3,4}$, Douglas J. Durian$^{3,1}$,\\
Kevin T. Turner$^{1,\ast}$\\
\normalsize{$^{1}$Department of Mechanical Engineering and Applied Mechanics,}\\
\normalsize{University of Pennsylvania, Philadelphia, PA 19104}\\
\normalsize{$^{2}$Department of Mechanical Engineering and Production,}\\
\normalsize{Aarhus University, Aarhus, Denmark}\\
\normalsize{$^{3}$Department of Physics and Astronomy,}\\ 
\normalsize{University of Pennsylvania, Philadelphia, PA 19104}\\
\normalsize{$^{4}$Department of Mechanical Engineering,}\\ 
\normalsize{University of Michigan, Ann Arbor, MI 48109}\\
\\
\normalsize{$^\ast$To whom correspondence should be addressed;} \\ 
\normalsize{E-Mails: fulco@seas.upenn.edu (S. Fulco), kturner@seas.upenn.edu (K.T. Turner).}
}

\date{}
\begin{document} 

%\baselineskip24pt

\maketitle 
\begin{sciabstract}
Mechanical metamaterials with engineered failure properties typically rely on periodic unit cell geometries or bespoke microstructures to achieve their unique properties. We demonstrate that intelligent use of disorder in metamaterials leads to distributed damage during failure, resulting in enhanced fracture toughness with minimal losses of strength. Toughness depends on the level of disorder, not a specific geometry, and the confined lattices studied exhibit a maximum toughness enhancement at an optimal level of disorder. A mechanics model that relates disorder to toughness without knowledge of the crack path is presented. The model is verified through finite element simulations and experiments utilizing photoelasticity to visualize damage during failure. At the optimal level of disorder, the toughness is more than 2.6$\times$ of an ordered lattice of equivalent density.
\end{sciabstract}
%%%%%%%%%%%%%%%%%%%%%%%%%%%%%%%%%%%%%%%%%%%%%%%%%%%%%%%%%%%%%%%%%%

% In setting up this template for *Science* papers, we've used both
% the \section* command and the \paragraph* command for topical
% divisions.  Which you use will of course depend on the type of paper
% you're writing.  Review Articles tend to have displayed headings, for
% which \section* is more appropriate; Research Articles, when they have
% formal topical divisions at all, tend to signal them with bold text
% that runs into the paragraph, for which \paragraph* is the right
% choice.  Either way, use the asterisk (*) modifier, as shown, to
\pagebreak
Architected mechanical metamaterials leverage geometry to achieve outstanding mechanical properties, including superior stiffness- and strength-to-weight ratios \cite{ Compton_AdvMater_2014,Schaedler_Science_2011,Mueller_AdvMater_2018,Zheng_Science_2014,Bauer_NatMat_2016,Meza_Science_2014,Park_ACSNano_2020,Ramachandramoorthy_AppMatToday_2020}. Only recently has the fracture behavior of architected materials been investigated \cite{Jorgensen_JMPS_2020,Shaikeea_NatureMaterials_2022,Mateos_AdvFuncMat_2019,Athanasiadis_EML_2021,Hossain_JMPS_2014,Hsueh_JourMechPhysSol_2018,Luan_JMPS_2022,OMasta_JMPS_2017,Zhang_NewJourPhys_2018,Cui_IJSS_2020,MuroBarrios_JMPS_2022,Fulco_EML_2022,Fulco_JMPS_2024}, despite the importance of failure via crack propagation in many materials. Architecture provides an opportunity to improve mechanical performance by controlling and enhancing fracture toughness, but this has not been fully realized, as previous investigations have primarily focused on materials with simple, periodic microstructures \cite{Hutchinson_JMPS_2005,Maiti_ScriptaMetal_1984, Fleck_JMPS_2007}.
\par Seminal works on the fracture of lattice materials \cite{Hutchinson_JMPS_2005,Maiti_ScriptaMetal_1984, Fleck_JMPS_2007, Luan_SciAdv_2023} showed that the toughness, $G_c$, is primarily a function of the connectivity (controlling if the lattice is stretch- or bending-dominated), the relative density $\rho^n$ ($n = 1$ if stretch-dominated and $n>1$ if bending-dominated), the unit cell size, $L$ \cite{Fleck_JMPS_2007}, and the orientation of the ligaments. Thus, for a fixed relative density, stretch-dominated lattices with large unit cells achieve the highest toughness, with the only other meaningful variable being the orientation of the ligaments, the effects of which are not well understood.
\par Although often inspired by nature \cite{ Barthelat_NatReviews_2016,Glassmaker_PNAS_2007,Bacca_JMPS_2016,Chung_JourRoySocInt_2005}, the structural complexity of architected materials is comparatively limited, predominantly utilizing regular, periodic patterns \cite{ Hossain_JMPS_2014,Jorgensen_JMPS_2020,Shaikeea_NatureMaterials_2022,Mateos_AdvFuncMat_2019,Athanasiadis_EML_2021}. This simplicity contrasts sharply with the disordered and non-periodic structures found in natural materials with high fracture toughness, such as bone \cite{Yang_Biomat_2006} and nacre \cite{Huang_SciRep_2013}. Natural structures are also often confined or a component in a larger system \cite{Barthelat_NatReviews_2016, Yang_Biomat_2006}, unlike many of the freestanding structures that have been explored to date \cite{Shaikeea_NatureMaterials_2022, Choukir_IJES_2023}. When more complicated or disordered geometries are investigated \cite{Choukir_IJES_2023}, typically a small number of representative geometries are considered, leaving it unclear if the measured properties are a product of the disorder and thus generalizable, or are unique to the specific geometry.
\par Disordered structures have not been systematically studied primarily because traditional fracture mechanics analyses of architected materials assume a self-similar crack front during propagation \cite{Hutchinson_JMPS_2005,Maiti_ScriptaMetal_1984, Fleck_JMPS_2007}, limiting the analysis to periodic structures typically at small length scales. However, these same analyses show that toughness is enhanced by increasing the unit cell size. Previous works \cite{Hedvard_IJSS_2024,Fulco_EML_2022,Fulco_JMPS_2024} on fracture of ordered architected materials posited that the largest structures that are expected to significantly influence the toughness have a comparable length scale to the process zone, $\lambda_0$ \cite{Irwin_JAM_1957}, the region of inelastic damage resulting from the singular crack-tip stresses. In an elastic-brittle lattice, the stresses are not truly singular, as the crack-tip is not sharp, nor is there a traditional process zone arising from inelastic effects. However, there remains a region of high stress near the crack tip and a length scale over which these stresses decay that is akin to an effective process zone size \cite{Hedvard_IJSS_2024}. Features larger than the decay length are unlikely to impact the toughness as the local fracture stresses will fully develop within the feature itself. 
\par While structures comparable to this length scale have already demonstrated significant enhancements in toughness \cite{Hedvard_IJSS_2024,Fulco_EML_2022,Fulco_JMPS_2024}, they remain periodic and assume an a priori crack path and a self-similar crack front. Identifying the crack path in complicated disordered structures is challenging with recent works using data-driven approaches, such as graph neural networks \cite{Karapiperis_ComEng_2023}, to predict crack path and toughness. These approaches, however, do not provide insight into the mechanisms through which disorder affects toughness.
\par We investigate the fracture of 2-D plane strain triangular lattices with unit cells comparable to the effective process zone in size, and quantify how geometric disorder relates to the distribution of damage during crack propagation and the toughness. We introduce a framework utilizing an effective crack length that demonstrates how toughness may be estimated by identifying the number of ligament failures per crack advance and the strengths of the lattice unit cells. Results are verified through a combination of finite element simulations and fracture experiments, which reveal that the introduction of disorder changes the damage from a straight, continuous crack path to discontinuous distributed damage, resulting in higher toughness. For a confined lattice, we identify an optimal level of disorder where toughness is maximized. These enhancements in toughness are achieved with minimal loss of strength and negligible change in stiffness. 
%%%%%%%%%%%%%%%%%%%%%%%%%%%%%%%%%%%%%%%%%%%%%%%%%%%%%%%%%%%%%%%%%%
%%%%%%%%%%%%%%%%%%%%%%%%%%%%%%%%%%%%%%%%%%%%%%%%%%%%%%%%%%%%%%%%%%
\begin{figure}[ht]
    \centering
    \includegraphics[width=\textwidth]{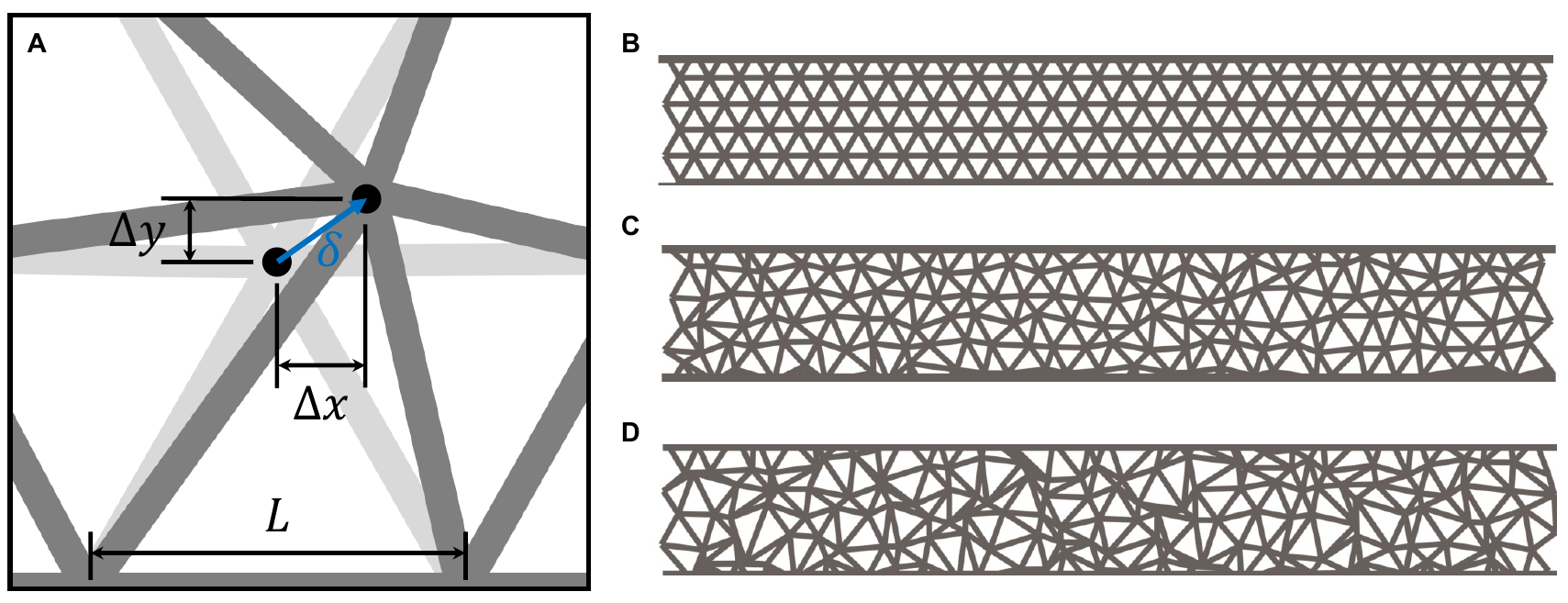}
    \caption{(A) A single unit cell showing the introduction of disorder by perturbing node locations by $\delta$. Representative networks with varying average levels of disorder (B) $\bar\delta = 0$, (C) $\bar\delta=0.15$, and (D) $\bar\delta=0.27$.}
    \label{fig:fig13}
\end{figure}
%%%%%%%%%%%%%%%%%%%%%%%%%%%%%%%%%%%%%%%%%%%%%%%%%%%%%%%%%%%%%%%%%%
%%%%%%%%%%%%%%%%%%%%%%%%%%%%%%%%%%%%%%%%%%%%%%%%%%%%%%%%%%%%%%%%%%
\par We consider a lattice of finite height (e.g., a lattice at an interface), as shown in Fig. \ref{fig:fig13}(B-D). Geometric disorder is incorporated while maintaining consistent connectivity by perturbing the node locations in 2-D by distances $\Delta x$ and $\Delta y$ (Fig. \ref{fig:fig13}(A)). Each perturbation is selected from a normal distribution. For a unit cell with length $L$, the disorder magnitude is defined as $\delta=\sqrt{\Delta x^2 + \Delta y^2}/L$. The network's average disorder, $\bar\delta$, is calculated as the mean of all the perturbations, with larger $\bar\delta$ values corresponding to greater disorder across the network, as shown in Figs. \ref{fig:fig13}(C) and (D). The impacts of different definitions of disorder are discussed below. Lattices with perturbations that result in overlapping nodes are not considered, as they would change the connectivity of the lattice, which is outside the scope of this analysis. This places a geometric limit on the perturbation of one node to be less than $\delta=0.433$. Additional practical constraints arising from the finite thickness of the beams and neighboring nodes limit the average level of disorder in a lattice to less than $\bar\delta\approx 0.375$.
%%%%%%%%%%%%%%%%%%%%%%%%%%%%%%%%%%%%%%%%%%%%%%%%%%%%%%%%%%%%%%%%%%
\section*{Effect of Disorder on Local Lattice Failure}
%%%%%%%%%%%%%%%%%%%%%%%%%%%%%%%%%%%%%%%%%%%%%%%%%%%%%%%%%%%%%%%%%%
\par The impact of disorder on the local strength of the lattice is assessed by a unit cell analysis consisting of a single perturbed node and its connecting ligaments, as shown in Fig. \ref{fig:fig13}(A). As the lattice is stretch-dominated, only the axial stresses in the ligaments are significant \cite{Ashby_PhilTransRoySoc_2006}. Bending stresses and stress concentrations at the nodes are neglected. The axial stress in each ligament is $\sigma_{a_i} = F_i / A$, where $i$ denotes the ligament, $F_i$ is the force in the ligament, and $A = tb$ is the ligament's cross-sectional area, with $t$ being the in-plane width and $b$ the out-of-plane thickness. An average normal stress $\bar\sigma_{yy}$ is applied to the upper and lower boundaries of the unit cell, and equilibrium at each node requires that $\sum\limits_{i\epsilon N_j} F_{i_{x}} = 0$ and $\sum\limits_{i\epsilon N_j} F_{i_{y}} = 0$, at each node $j$, where $N_j$ are the set of ligaments that connect to each node. The solutions of these equations determine all $\sigma_{a_i}$. Assuming the ligaments are elastic-brittle with a failure stress, $\sigma_f$, this model predicts the unit cell's strength, $\sigma_m$, for any node perturbation. This approach applies to stretch-dominated lattices with narrow ligaments ($t = L/10$ here), but can be adapted to other cases through more sophisticated failure criteria. 
%%%%%%%%%%%%%%%%%%%%%%%%%%%%%%%%%%%%%%%%%%%%%%%%%%%%%%%%%%%%%%%%%%
%%%%%%%%%%%%%%%%%%%%%%%%%%%%%%%%%%%%%%%%%%%%%%%%%%%%%%%%%%%%%%%%%%
\begin{figure}[ht]
    \centering
    \includegraphics[width=\textwidth]{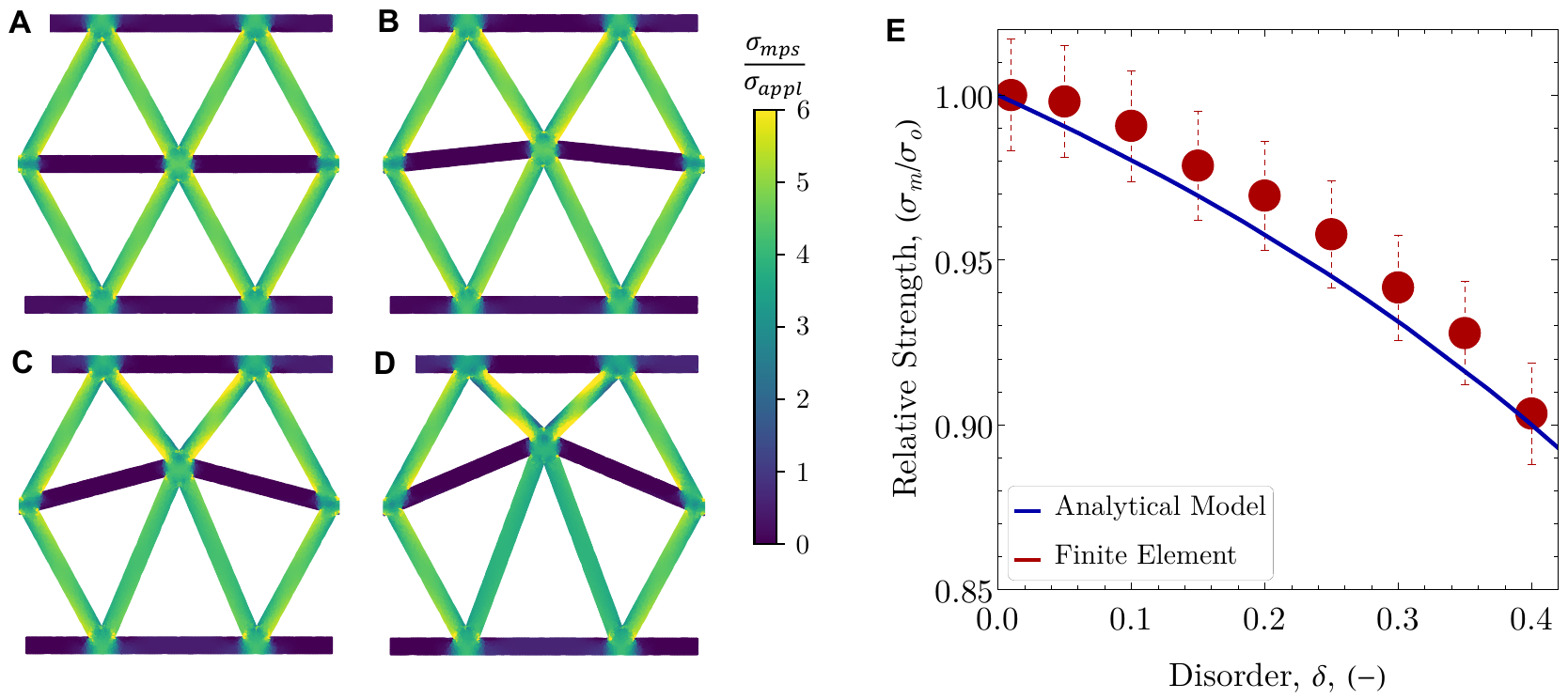}
    \caption{Representative finite element stress distributions showing the maximum principal stresses $\sigma_{mps}$ relative to the average applied stress, $\sigma_{appl}$, for unit cells with (A) $\delta=0$, (B) $\delta=0.10$, (C) $\delta=0.20$, (D) $\delta=0.30$. (E) Relative unit cell strength as a function of disorder. Result from the analytical model is also indicated by the solid curve.}
    \label{fig:fig0}
\end{figure}
%%%%%%%%%%%%%%%%%%%%%%%%%%%%%%%%%%%%%%%%%%%%%%%%%%%%%%%%%%%%%%%%%%
%%%%%%%%%%%%%%%%%%%%%%%%%%%%%%%%%%%%%%%%%%%%%%%%%%%%%%%%%%%%%%%%%%
\par The predictions from the analytical model are verified using 2-D finite element simulations of the unit cell with varying perturbations to the central node (Fig. \ref{fig:fig0})(A-D) under uniaxial tension ($\bar\sigma_{yy}=\sigma_{appl}$). The node is perturbed at disorder $\delta$ from 0 to 0.4 in increments of 0.05, with seven simulations at each level of disorder corresponding to perturbing the node an angle from 0$\degree$ to 180$\degree$ in increments of 30$\degree$. Periodic boundary conditions were applied to the left and right sides of the unit cell. The average maximum principal stress, $\sigma_{mps}$, in the middle of each ligament was used to predict the failure.
\par The strengths of the unit cells as a function of disorder, relative to an unperturbed unit cell are given in Fig. \ref{fig:fig0}(E) for the simulations and the analytical model. The agreement between the two is close, although the analytical model slightly underestimates the strength. This discrepancy arises primarily from the omission of periodic boundary conditions in the analytical model. The strength of the unit cell monotonically decreases with $\delta$, but the loss of strength remains $\leq10\%$ on average at the highest level of disorder. The reduction in strength is due to the loss of symmetry, with the stress being transferred to at least one ligament that is oriented more along the line of loading than the others, causing premature failure.
%%%%%%%%%%%%%%%%%%%%%%%%%%%%%%%%%%%%%%%%%%%%%%%%%%%%%%%%%%%%%%%%%%
\section*{Effect of Disorder on Toughness}
%%%%%%%%%%%%%%%%%%%%%%%%%%%%%%%%%%%%%%%%%%%%%%%%%%%%%%%%%%%%%%%%%%
\par The fracture behaviors of lattices with varying geometries were calculated using 2-D linear elastic finite element simulations. Crack growth was simulated by running the model multiple times, as described below.  For a given network geometry, the model was run once to calculate stresses in the network. The ligament with the highest stress was identified and deleted. The updated model was then run again to identify the next ligament that would fail. This process was repeated for up to 50 ligament failures. As the applied load and stresses are linearly related, the applied load at each ligament failure was determined by scaling the ligament stress to the failure stress. Additional details about the model geometry and simulations are provided in the Materials and Methods section.
\par
\begin{figure}[ht]
    \centering
    \includegraphics[width=\textwidth]{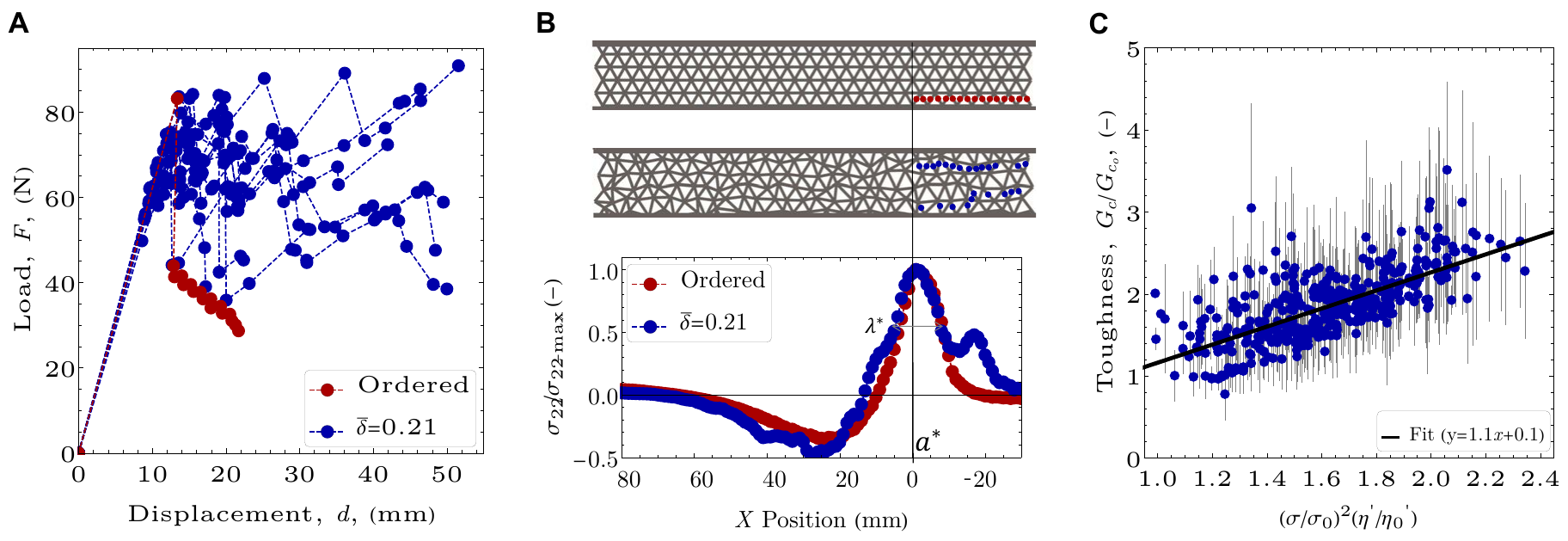}
    \caption{(A) Representative load-displacement behavior for the ordered lattice and ten geometries with $\bar\delta=0.15$. (B) Representative ordered and disordered geometries, with ligament failure indicated. The average stress profiles along the crack plane are shown with the effective crack tip at the peak stress indicated along with the process zone, $\lambda^\ast$. (C) Toughness (ave. $\pm$ std. dev. for all ligament ruptures) of each simulated geometry, relative to the ordered lattice. Results plotted versus eq. (\ref{eq:GcGo}). A linear fit ($R^2=0.97$) is also indicated}
    \label{fig:fig23}
\end{figure}

 The load-displacement curves for the ordered lattice and ten distinct disordered lattices, each with $\bar\delta \approx 0.15$, are shown in Fig. \ref{fig:fig23}(A). While the damage curves for the disordered configurations exhibit considerable variability, they predominantly maintain loads that exceed those of the ordered lattice, indicating higher damage resistance. Crack propagation and toughness were investigated for 420 unique geometries of varying levels of disorder between $0 \leq \bar\delta \leq 0.375$. This allows for the effect of disorder to be quantified independent of the specific lattice geometry. 
\par The crack path through the network, as shown for the representative geometries in Fig. \ref{fig:fig23}(B), is identified by tracking the location of ligament failures. In the case of the ordered lattice, crack propagation follows a straight path, akin to a crack in a homogeneous material. Conversely, in most disordered lattices, the damage path is neither straight nor always continuous. We define an effective crack tip by considering the stress distribution in the network. As shown in Fig. \ref{fig:fig23}(B), averaging the stress across the height of the network at every $x$-location, results in a homogenized stress profile that is similar to the stress distribution around a crack tip in a continuum material with a process zone \cite{Irwin_JAM_1957}. The effective crack length, $a^\ast$, is determined by identifying the location of the maximum tensile stress arising from the singular-like region of the stress distribution (Fig. \ref{fig:fig23}(B)). This methodology for defining $a^\ast$ not only aligns with the physical behavior of cracks but also simplifies the subsequent toughness calculation, since the effective crack length monotonically increases during failure but does not require an assumption of self-similarity of the crack front. The stress distribution also provides a measurement of the effective process zone size. We define the effective process zone size, $\lambda^\ast$, as the full-width at half-maximum of the stress distribution around the effective crack length. For the ordered lattice, $\lambda^\ast\approx 2L$, which confirms that the unit cell is comparable in size to the process zone. 
%%%%%%%%%%%%%%%%%%%%%%%%%%%%%%%%%%%%%%%%%%%%%%%%%%%%%%%%%%%%%%%%%%
%%%%%%%%%%%%%%%%%%%%%%%%%%%%%%%%%%%%%%%%%%%%%%%%%%%%%%%%%%%%%%%%%%
%%%%%%%%%%%%%%%%%%%%%%%%%%%%%%%%%%%%%%%%%%%%%%%%%%%%%%%%%%%%%%%%%%
%%%%%%%%%%%%%%%%%%%%%%%%%%%%%%%%%%%%%%%%%%%%%%%%%%%%%%%%%%%%%%%%%%
\par The toughness is predicted by considering the strain energy change during crack propagation. Most of the strain energy change is localized to the unit cells around the ligaments that rupture, giving the total change in strain energy, $\Delta U_{el}$, released during crack propagation as
%%%%%%%%%%%%%%%%%%%%%%%%%%%%%%
%%%%%%%%%%%%%%%%%%%%%%%%%%%%%%
\begin{equation}
   \Delta U_{el}=\frac{\eta}{2}\int_V \epsilon \sigma_m dV,
\end{equation}
%%%%%%%%%%%%%%%%%%%%%%%%%%%%%%
%%%%%%%%%%%%%%%%%%%%%%%%%%%%%%
where $V$ is the volume of the ligaments in the unit cell, $\sigma_m$ is the stress in the unit cell at failure, $\epsilon$ is the strain, and $\eta$ is the number of ligaments that fail. Since the network is stretch-dominated, stress and strain are taken as approximately uniform through the volume of each ligament, with $\epsilon_m = \sigma_m/E$, where $E$ is the Young's modulus. Thus, the critical energy release rate of the 2-D lattice is
%%%%%%%%%%%%%%%%%%%%%%%%%%%%%%
%%%%%%%%%%%%%%%%%%%%%%%%%%%%%%
\begin{equation}
    G_c=-\frac{1}{b}\frac{\Delta U}{\Delta a^\ast}\approx-\frac{1}{b}\left(\frac{\Bar{\sigma}_m^2}{E}\right)V\eta^{'},
\end{equation}
%%%%%%%%%%%%%%%%%%%%%%%%%%%%%%
%%%%%%%%%%%%%%%%%%%%%%%%%%%%%%
where $\eta^{'}=\Delta\eta/\Delta a^{\ast}$, and $\Bar{\sigma}_m$ is the average stress in the unit cell at failure. Thus, the relative toughness of a disordered lattice, compared to the ordered lattice toughness, $G_o$, is
%%%%%%%%%%%%%%%%%%%%%%%%%%%%%%
%%%%%%%%%%%%%%%%%%%%%%%%%%%%%%
\begin{equation}
    \frac{G_c}{G_o}=\left(\frac{\bar\sigma_m}{\bar\sigma_0}\right)^2 \left(\frac{\eta^{'}}{\eta_o^{'}}\right),
    \label{eq:GcGo}
\end{equation}
%%%%%%%%%%%%%%%%%%%%%%%%%%%%%%
%%%%%%%%%%%%%%%%%%%%%%%%%%%%%%
where $\bar\sigma_o$ is the strength of an unperturbed unit cell, $\eta_o^{'}$ is the number of ligament failures per effective crack propagation for the ordered lattice, and any small volume changes of the ligaments between the ordered and disordered lattices are neglected. 
\par The first term on the right-hand side of eq. (\ref{eq:GcGo}) is $\leq 1$ for all $\delta$ (see Fig. \ref{fig:fig0}(E)). Thus, any enhancements in toughness must come from sufficient distributed damage to account for the loss in local strength. The toughness of all simulated geometries (mean $\pm$ std. dev.), relative to the ordered lattice, are shown in Fig. \ref{fig:fig23}(C) as a function of eq. (\ref{eq:GcGo}). The results are well-described by the relationship and significant enhancements in toughness (up to more than $3\times$) are achieved as a result of the distributed damage. A linear fit to the data in Fig. 3(C) ($R^2=0.97$) has a slope of 1.1, slightly higher than the predicted slope of 1 due to a number of outlier cases that are high and none that are low. This is expected based on eq. (\ref{eq:GcGo}), as there is no mechanism to generate significantly lower toughness than the ordered lattice. The number of ligament failures cannot decrease relative to the ordered lattice, and the losses in local strength are small ($\leq 10\%$), resulting in a toughness that is typically equal to, or greater than, the ordered lattice.
%%%%%%%%%%%%%%%%%%%%%%%%%%%%%%%%%%%%%%%%%%%%%%%%%%%%%%%%%%%%%%%%%%
\section*{Experimental Results \& Discussion}
%%%%%%%%%%%%%%%%%%%%%%%%%%%%%%%%%%%%%%%%%%%%%%%%%%%%%%%%%%%%%%%%%%
\par Fracture experiments were performed on laser-cut polymethylmethacrylate (PMMA) specimens as detailed in the Materials and Methods section \cite{Ute_Polymer_1995}. Three sets of specimens were prepared: an ordered lattice, and two sets of disordered structures. The disordered structures were generated by selecting two geometries randomly from the simulated lattices, and then scaling the average perturbation of each geometry to achieve average levels of disorder in the range $0\leq \bar\delta \leq 0.34$. This allowed the effect of the disorder magnitude to be investigated without introducing any bias arising from specific geometries. A total of 33 specimens were tested, which included 3 ordered specimens, and 15 specimens in each of the disordered sets.
\par During fracture testing, specimens were analyzed using photoelasticity. For a photoelastic material such as PMMA \cite{Daniels_RevSciInst_2017}, the local stresses in the material alter the propagation of light through the specimen. When imaged through cross polarizers, this effect is visible such that the intensity of the observed light is directly related to the difference in principal stresses.
%%%%%%%%%%%%%%%%%%%%%%%%%%%%%%%%%%%%%%%%%%%%%%%%%%%%%%%%%%%%%%%%%%
%%%%%%%%%%%%%%%%%%%%%%%%%%%%%%%%%%%%%%%%%%%%%%%%%%%%%%%%%%%%%%%%%%
\begin{figure}[ht]
    \centering
    \includegraphics[width=\textwidth]{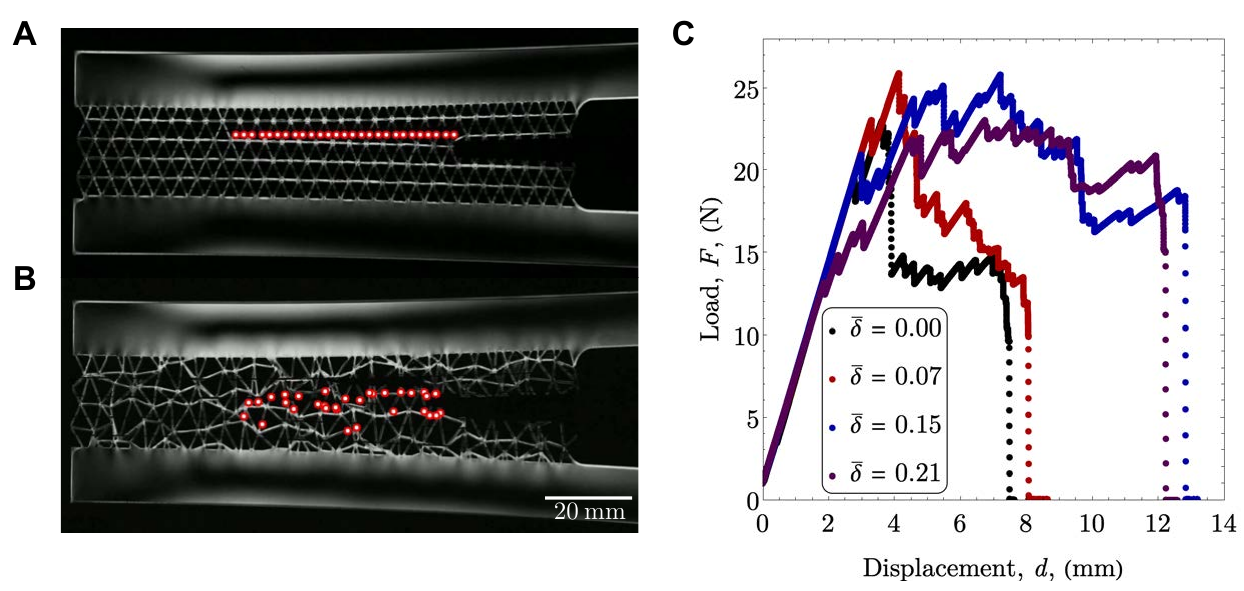}
    \caption{Photoelastic images of (A) an ordered lattice, and (B) a disordered lattice ($\bar\delta=0.27$), with ligament failures indicated. (C) Representative load-displacement behavior for experimental specimens at varying levels of disorder.}
    \label{fig:fig33}
\end{figure}
%%%%%%%%%%%%%%%%%%%%%%%%%%%%%%%%%%%%%%%%%%%%%%%%%%%%%%%%%%%%%%%%%%
%%%%%%%%%%%%%%%%%%%%%%%%%%%%%%%%%%%%%%%%%%%%%%%%%%%%%%%%%%%%%%%%%% 
This technique allows for the identification of failing ligaments by monitoring changes in local light intensity during fracture. Instances of ligament failure are shown in Fig. \ref{fig:fig33}(A) for an ordered lattice and in (B) for a disordered lattice, where each specimen has the same length of crack propagation. The observations reveal that disordered lattices exhibit more distributed damage and a higher number of ligament failures during fracture (33 for a disordered lattice with $\bar\delta=0.27$ vs. 27 for the ordered lattice for cracks propagating over ${\sim}100$ mm). 
\par Representative load-displacement curves for ordered and disordered lattices are given in Fig. \ref{fig:fig33}(C). Specimens with disordered structures sustained higher loads during failure compared to the ordered lattice. However, specimens at the highest levels of disorder ($\delta \geq 15\%$) showed no additional enhancements in load capacity.
%%%%%%%%%%%%%%%%%%%%%%%%%%%%%%%%%%%%%%%%%%%%%%%%%%%%%%%%%%%%%%%%%%
%%%%%%%%%%%%%%%%%%%%%%%%%%%%%%%%%%%%%%%%%%%%%%%%%%%%%%%%%%%%%%%%%%
\begin{figure}[ht]
    \centering
    \includegraphics[width=0.5\textwidth]{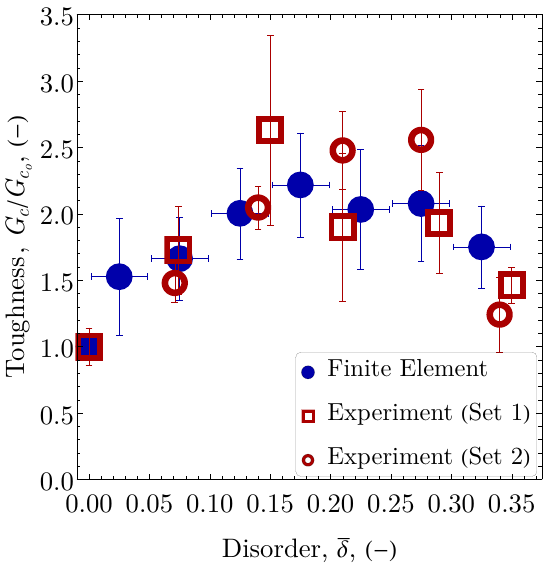}
    \caption{(A) Toughness, relative to the ordered lattice, as a function of disorder percentage, for all finite element simulations (mean $\pm$ std. dev. of $\sim 60$ geometries for each point), and experiments (mean $\pm$ std. dev. for 3 identical specimens at each point).}
    \label{fig:fig43}
\end{figure}
%%%%%%%%%%%%%%%%%%%%%%%%%%%%%%%%%%%%%%%%%%%%%%%%%%%%%%%%%%%%%%%%%%
%%%%%%%%%%%%%%%%%%%%%%%%%%%%%%%%%%%%%%%%%%%%%%%%%%%%%%%%%%%%%%%%%%
\par Experimental toughness values as a function of the disorder are shown in Fig. \ref{fig:fig43}, with each point being the mean and standard deviations for three specimens. The toughness of the lattices was calculated from the load-displacement data using a modified double cantilever beam analysis \cite{Olsson_CompSciTech_1992}, as described in the Materials and Methods section. Toughness predictions from the finite element simulations are also shown, where each data point corresponds to ${\sim}60$ geometries (mean $\pm$ std. dev.). By considering many geometries at each level of disorder, the toughness represents the average  toughness for a system where the number of unit cells is large, rather than simply a measurement of any one unique geometry. The qualitative and quantitative agreement between the simulations and experiments further verifies this result.
\par Consistent with the predictions of eq. (\ref{eq:GcGo}), distributed damage enhances the toughness of the structure compared to the ordered lattice, since $\eta^{'}/\eta_o^{'}>1$. The effects of confinement place an inherent constraint on the extent of the distributed damage and, consequently, the maximum possible toughness enhancement. When the peak value of $\eta^{'}$ is reached but disorder increases further, a reduction in toughness is expected due to diminished local strength. This is apparent in both the experimental and simulation data, where the introduction of disorder initially leads to a rapid increase in toughness as a result of the distributed damage. The enhancement plateaus at $\bar\delta=0.15$ due to the confined nature of the lattice studied here, and beyond $\bar\delta=0.2$ the toughness begins to decrease as the local strength is reduced. 
\par Experimentally, the maximum toughness achieved was 2.6$\times$ that of the ordered lattice at $\bar\delta=0.15$. The value of the optimal level of disorder is a function of the definition of disorder, which is not standardized. Other methods for characterizing disorder, such as Voronoi tessellation \cite{Zhu_JMPS_2001}, will provide qualitatively identical results to those shown here, but at varying quantitative levels of disorder.

%%%%%%%%%%%%%%%%%%%%%%%%%%%%%%%%%%%%%%%%%%%%%%%%%%%%%%%%%%%%%%%%%%
\section*{Conclusions}
%%%%%%%%%%%%%%%%%%%%%%%%%%%%%%%%%%%%%%%%%%%%%%%%%%%%%%%%%%%%%%%%%%
\par Disorder increases the fracture toughness of architected lattice materials by generating distributed damage. Through finite element simulations of varying lattice geometries, enhancements are shown to be common to the set of geometries characterized by the level of disorder, not for any specific architecture. These enhancements are achieved with minimal losses in strength of $\leq 10\%$, relative to the ordered lattice, and negligible changes in stiffness. For a confined lattice, a maximum enhancement in toughness is identified due to the inherent limit of the extent of the distributed damage, leading to an optimal level of disorder. The enhancements in toughness are verified through experiments using photoelasticity to visualize the damage during failure. The maximum increase in toughness was 2.6$\times$ at $\bar\delta=0.15$. These enhancements in toughness demonstrate how constraining the consideration of architected materials to periodic structures significantly limits the possible mechanical performance. 
\par A triangular lattice was considered in this work as a representative geometry, but the approach can be extended to a broad set of lattices with varying topology and connectivity. The approach is material-agnostic and is expected to apply to most elastic-brittle materials. The framework presented in this work can be applied to inform the failure analysis of a variety of disordered structures, significantly expanding the design spectrum of architected materials and enabling new enhancements in toughness.  

\section*{Materials and Methods}
%%%%%%%%%%%%%%%%%%%%%%%%%%%%%%%%%%%%%%%%%%%%%%%%%%%%%%%%%%%%%%%%%%
\subsection*{Specimen Fabrication}
%%%%%%%%%%%%%%%%%%%%%%%%%%%%%%%%%%%%%%%%%%%%%%%%%%%%%%%%%%%%%%%%%%
Fracture specimens were fabricated from 6 mm-thick, cast, transparent polymethylmethacrylate (PMMA) sheets, which were laser-cut using an x1250 laser cutter (Eduard, Inc., Denmark) with a 150-Watt laser operating at 80\% power and a cutting speed of 15 mm/min. A dual-pass laser cutting approach was used to mitigate the effects of residual heat. Furthermore, to avoid unintended failure of the solid beams and to isolate failure to the lattice ligaments, rather than the nodes, an additional raster cut was introduced in the middle of each ligament at a laser speed of 30 mm/min, resulting in a consistent cut depth of  ${\sim} 2$ mm. 
%%%%%%%%%%%%%%%%%%%%%%%%%%%%%%%%%%%%%%%%%%%%%%%%%%%%%%%%%%%%%%%%%%
%%%%%%%%%%%%%%%%%%%%%%%%%%%%%%%%%%%%%%%%%%%%%%%%%%%%%%%%%%%%%%%%%%
\begin{figure}[ht]
    \centering
    \includegraphics[width=0.5\textwidth]{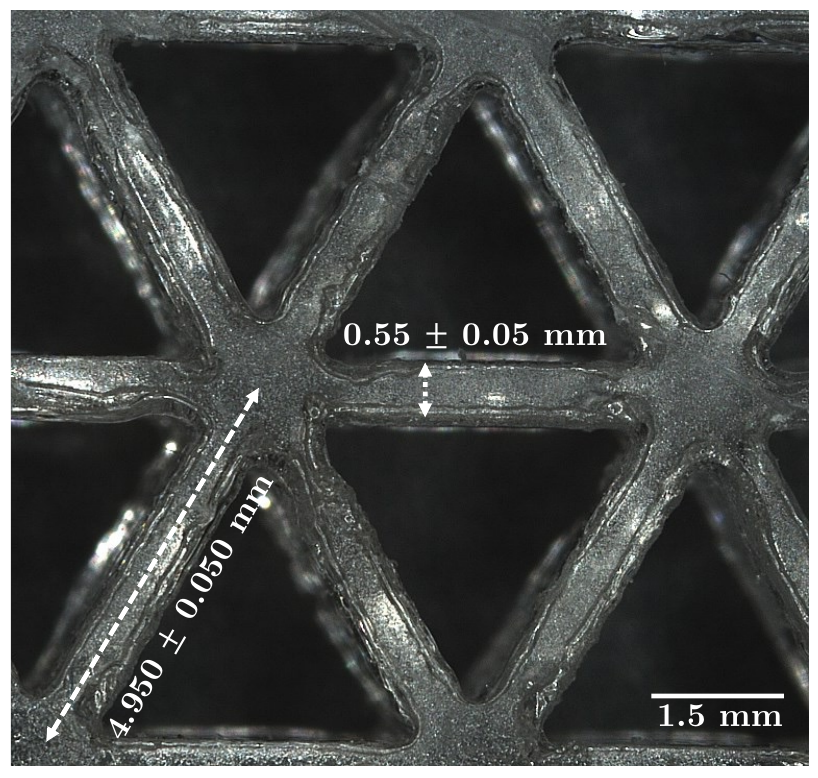}
    \caption{Optical image of laser-cut PMMA lattice (without raster cuts) with ligament width and unit cell length indicated}
    \label{fig:figS1_3}
\end{figure}
%%%%%%%%%%%%%%%%%%%%%%%%%%%%%%%%%%%%%%%%%%%%%%%%%%%%%%%%%%%%%%%%%%
%%%%%%%%%%%%%%%%%%%%%%%%%%%%%%%%%%%%%%%%%%%%%%%%%%%%%%%%%%%%%%%%%%
\par Ligament thicknesses were measured optically and found to be $t=$ 0.55 mm $\pm$ 0.05 mm, as shown in Fig. \ref{fig:figS1_3}(A), and the unit cell length was measured to be $L=$ 4.95 mm $\pm$ 0.05, both of which closely match the nominal value used in the simulations. Specimens were annealed for 4 hours in a 90\degree C oven to relieve residual stresses from fabrication. This temperature was chosen as it is close to but below the glass transition temperature of PMMA (\textit{37}). 
\par
\begin{figure}[ht]
    \centering
    \includegraphics[width=\textwidth]{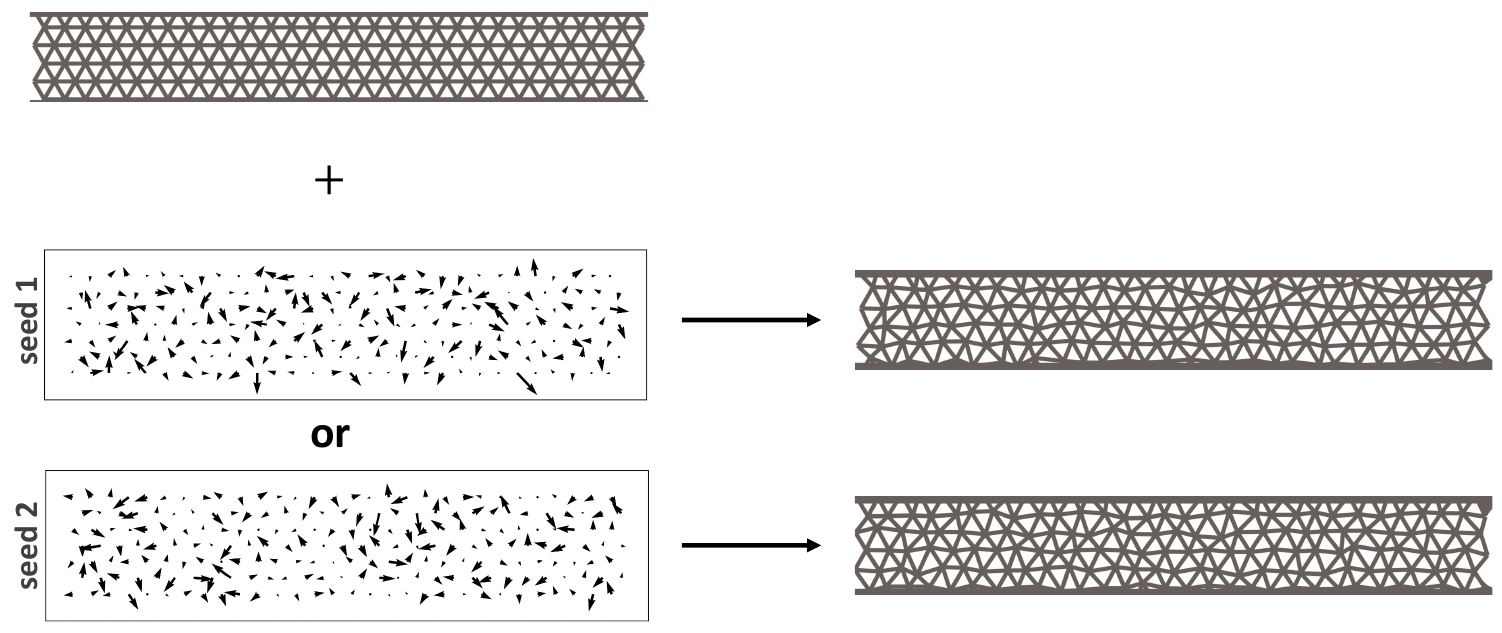}
    \caption{Representative disordered lattices used for the experiments, with $\bar\delta\approx 0.15$ in both cases. To generate the disordered lattices, an ordered lattice is perturbed following the directions and relative magnitudes shown for each seed. To generate lattices of varying levels of disorder, the average magnitude of the perturbations is changed.}
    \label{fig:figS3_3}
\end{figure}

 Two sets of disordered lattices were fabricated using the seeds shown in Fig. \ref{fig:figS3_3}. To generate disordered lattices, the nodes of the ordered lattice are perturbed as shown. Lattices of varying magnitudes of disorder are generated by changing the average magnitude of the perturbations, with the directions and relative levels of perturbation being conserved. Structures at five levels of disorder between $\bar\delta=0.07 – 0.34$ are generated, along with an ordered lattice with $\bar\delta=0$ for a total of 11 unique geometries. Three identical specimens of each geometry level were fabricated and tested (33 total specimens).
%%%%%%%%%%%%%%%%%%%%%%%%%%%%%%%%%%%%%%%%%%%%%%%%%%%%%%%%%%%%%%%%%%
%%%%%%%%%%%%%%%%%%%%%%%%%%%%%%%%%%%%%%%%%%%%%%%%%%%%%%%%%%%%%%%%%%
%%%%%%%%%%%%%%%%%%%%%%%%%%%%%%%%%%%%%%%%%%%%%%%%%%%%%%%%%%%%%%%%%%
%%%%%%%%%%%%%%%%%%%%%%%%%%%%%%%%%%%%%%%%%%%%%%%%%%%%%%%%%%%%%%%%%%
%%%%%%%%%%%%%%%%%%%%%%%%%%%%%%%%%%%%%%%%%%%%%%%%%%%%%%%%%%%%%%%%%%
\subsection*{Fracture Testing \& Photoelasticity}
%%%%%%%%%%%%%%%%%%%%%%%%%%%%%%%%%%%%%%%%%%%%%%%%%%%%%%%%%%%%%%%%%%
\par Specimens were tested on a custom-built tensile testing instrument with dual-head control. Specimens were tested under displacement control, with each head moving at a rate of 2 mm/min, for a total displacement rate of 4 mm/min. Load and displacement measurements were made at a rate of 100 Hz. Tests were terminated after a crack had propagated through the entire specimen. 
\par For each seed geometry and disorder level, one specimen was selected for imaging via photoelasticity. The uniformity across specimens enabled the use of a single specimen per group for representative analysis. The transmission photoelastic setup was implemented as detailed in (\textit{38}), incorporating circular polarizers (Edmund Optics, NJ) and employing white light illumination. Imaging was executed with a Point Grey camera, capturing data at a frequency of 2 Hz, to visualize the stress distribution and crack propagation behavior. The intensity of the observed light $I$, is known to be
%%%%%%%%%%%%%%%%%%%%%%%%%%%%%%
%%%%%%%%%%%%%%%%%%%%%%%%%%%%%%
\begin{equation}
I\propto \sin{\left(\frac{\pi C_\Lambda b\Delta\sigma}{\Lambda}\right)}^2,
\end{equation}
%%%%%%%%%%%%%%%%%%%%%%%%%%%%%%
%%%%%%%%%%%%%%%%%%%%%%%%%%%%%%
where $\Lambda$ is the wavelength of the light, $b$ is the specimen thickness, $\Delta\sigma$ is the difference in principal stresses, and $C_\Lambda$ is a material constant that relates the wavelength of the light to the photoelasticity of the material (\textit{38}).For a stretch-dominated lattice, the axial stress in the ligament is dominant, thus $\Delta\sigma \approx \sigma_a$. Thus, as ligaments fail, the intensity of the observed light in each ligament will change according to the resulting stress distribution. This change in light intensity is used to identify ligament failures and the path of damage through the lattice during fracture. 
\par Toughness was calculated from the load-displacement data using a modified analysis for a double cantilever beam, with corrections for shear and a compliant foundation using the Winkler foundation model. The crack length, $a$ was determined using the theoretical compliance (\textit{39}) of the system, $C$, given by 
%%%%%%%%%%%%%%%%%%%%%%%%%%%%%%
%%%%%%%%%%%%%%%%%%%%%%%%%%%%%%
\begin{equation}
    C=\frac{8}{Eb}\left[\frac{3}{2}\alpha^{0.75}+\left(\frac{E}{4K\mu}+3\sqrt{\alpha}\right)\left(\frac{a}{h}\right)+
    \left(3\alpha^{0.25}+\frac{3}{2\pi}\sqrt{\frac{E}{\mu}}\right)\left(\frac{a}{h}\right)^2+\left(\frac{a}{h}\right)^3\right],
\end{equation}
%%%%%%%%%%%%%%%%%%%%%%%%%%%%%%
%%%%%%%%%%%%%%%%%%%%%%%%%%%%%%
with $\alpha=E/6E_T$, where $E$ is the Young's modulus of the PMMA, $\mu$ is the shear modulus, here taken as $\mu=E/2(1+\nu)$ for an isotropic material, where $\nu\approx0.3$ is the Poisson's ratio, $E_T$ is the effective tensile modulus of the lattice, $K$ is a constant equal to $5/6$, and $h$ is the thickness of the beams.
\par In order to predict the effective crack length, the effective tensile modulus of the foundation, $E_T$ must be known. For a stretch-dominated lattice, the effective modulus is predicted (\textit{36}) to be $E_T\approx \frac{1}{3}(\rho)E$, where $\rho$, is the relative density of the lattice. For the lattices used in this work where $t=L/10$, the effective modulus is predicted to be $E_T\approx 0.129E$. This was found to be in good agreement with the experimentally measured compliance.
\par The energy release rate of the lattice can therefore be calculated as a function of crack length as
%%%%%%%%%%%%%%%%%%%%%%%%%%%%%%
%%%%%%%%%%%%%%%%%%%%%%%%%%%%%%
\begin{equation}
    G=\frac{P^2}{2b}\frac{\partial C}{\partial a},
\end{equation}
%%%%%%%%%%%%%%%%%%%%%%%%%%%%%%
%%%%%%%%%%%%%%%%%%%%%%%%%%%%%%
where $P$ is the load. For a homogeneous linear elastic material, the critical energy release rate, $G_c$, is expected to be a constant with continuous crack propagation. However, in the disordered lattices this is not guaranteed. Variations in toughness result in cracks that arrest and propagate, as shown by the jumps in the load-displacement curves in Fig. 4(C). The toughness of the lattice corresponds to the critical energy release rate such that crack advance occurs. Thus, the average toughness of the lattice was calculated from all $G$ values during which the crack length increased as indicated by the change in compliance of the specimen. 
%%%%%%%%%%%%%%%%%%%%%%%%%%%%%%%%%%%%%%%%%%%%%%%%%%%%%%%%%%%%%%%%%%
\subsection*{Model Generation \& Finite Element Simulations}
%%%%%%%%%%%%%%%%%%%%%%%%%%%%%%%%%%%%%%%%%%%%%%%%%%%%%%%%%%%%%%%%%%
\par
\begin{figure}[ht]
    \centering
    \includegraphics[width=0.5\textwidth]{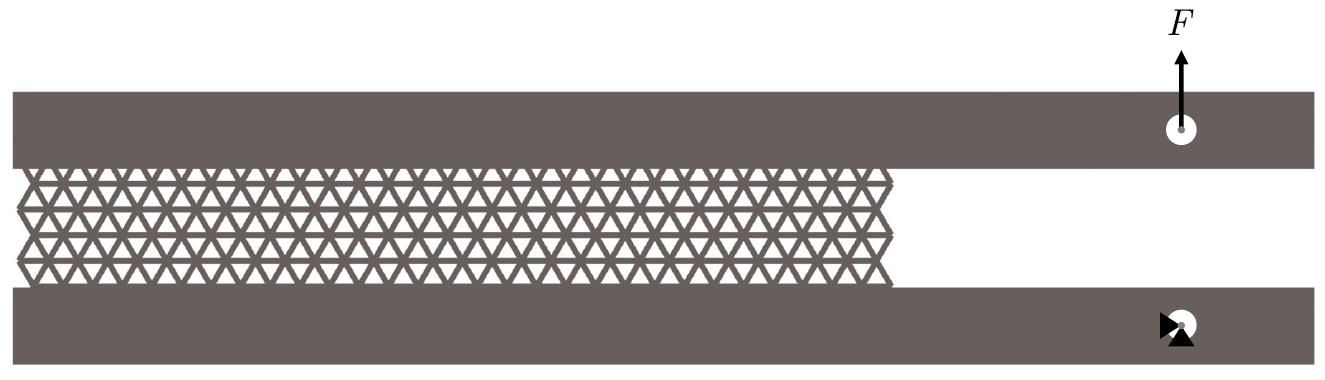}
    \caption{Double cantilever beam specimen geometry used for simulations and experiments, with an ordered lattice. The lower loading point is fixed horizontally and vertically, and the top loading point is displaced vertically, causing a vertical reaction force, $F$.}
    \label{fig:figS2_3}
\end{figure}

 Models were generated for finite element simulations using the open-source computer-aided design software, FreeCAD, integrated with Python. Lattice geometries were generated by creating a triangular lattice of nodes and connecting them with rectangular ligaments. To introduce disorder to the lattice, nodes were perturbed in $x$ and $y$ by selecting two random numbers from a Gaussian distribution with zero mean. As the magnitude of the variance in the distribution is increased, larger node perturbations occur. Using the ``random'' command from SciPy, pseudo-random numbers are generated instead of purely random numbers, with the exact number being reproducible if the same ``seed'' value is input into the number generator. For this work, ${\sim} 60$ seeds were generated, and scaled to 7 different levels of disorder (as shown in Fig. 7), creating 420 unique lattices. The lattices consist of 30 horizontal nodes and 5 vertical nodes, with a unit cell size of $L=5$ mm and ligament thickness $t=0.5$ mm. The overall length of the lattice is 150 mm, and the thickness is 20 mm, with a portion of the top and bottom of the lattice being embedded in the solid beams. The beams are 13 mm thick and 225 mm long, with the loading points 175 mm from the far edge of the lattice. A representative model geometry with an ordered lattice is shown in Fig. \ref{fig:figS2_3}, with the boundary conditions indicated.
%%%%%%%%%%%%%%%%%%%%%%%%%%%%%%%%%%%%%%%%%%%%%%%%%%%%%%%%%%%%%%%%%%
%%%%%%%%%%%%%%%%%%%%%%%%%%%%%%%%%%%%%%%%%%%%%%%%%%%%%%%%%%%%%%%%%%
%%%%%%%%%%%%%%%%%%%%%%%%%%%%%%%%%%%%%%%%%%%%%%%%%%%%%%%%%%%%%%%%%%
%%%%%%%%%%%%%%%%%%%%%%%%%%%%%%%%%%%%%%%%%%%%%%%%%%%%%%%%%%%%%%%%%%
\par Finite element simulations were performed in ABAQUS (2020, Providence, RI) by importing the FreeCAD model file of the lattice. The lattice was modeled under 2-D plane strain conditions, with one mounting point in the beams being fixed in $x$ and $y$, and the other under a 1 mm displacement boundary condition in $y$ and fixed in $x$. The elastic modulus was taken as $E=1$ GPa, and the failure stress was taken as $\sigma_f=20$ MPa. The linearity of the system allows for the results to be easily scaled to materials of differing stiffness and/or strength. The models were meshed with bilinear plane-strain quadrilateral elements (CPE4), with a free mesh in the lattice and a uniform mesh in the beams, with a mesh size of 0.05 mm. Approximately 110,000-130,000 elements were used per simulation, depending on the exact geometry. The mesh size was refined until there was a less than 0.5\% difference in the failure force of an ordered and representative disordered lattice with a 10\% increase in mesh density. 
\par To simulate crack propagation, the ligament under highest stress is identified in the finite element model. The applied displacement and corresponding load are scaled until the average von Mises stress in the ligament equals the failure stress, $\sigma_f$. The von Mises stress was used to account for possible nonuniformity in the stress distribution; however, as shown by the unit cell analysis, the results are expected to be very similar to a maximum principal stress criterion. The stresses are averaged in each ligament inside an area centered in the ligament with a radius of $2t$. This ligament is then deleted from the FreeCAD model and a new finite element simulation is run. This process is run multiple times and up to 50 times per geometry. Since each simulation in the sequence is treated as independent of the others, it is possible that the predicted displacement that a ligament fails at is smaller than the displacement at which the previous ligament failed. During a continuous fracture test under displacement control, this is not possible. Instead, these ligaments are expected to fail unstably during testing, and these data points are not considered in the analysis.

%%%%%%%%%%%%%%%%%%%%%%%%%%%%%%%%%%%%%%%%%%%%%%%%%%%%%%%%%%%%%%%%%%
\section*{Acknowledgments}
%%%%%%%%%%%%%%%%%%%%%%%%%%%%%%%%%%%%%%%%%%%%%%%%%%%%%%%%%%%%%%%%%%
This research was funded primarily by the National Science Foundation (NSF) MRSEC program [awards DMR-1720530 and DMR-2309043]. SF acknowledges support from the US Department of Defense (DoD) through the National Defense Science \& Engineering Graduate (NDSEG) Fellowship Program. MKB acknowledges the Villum Foundations for support under the Villum Experiment programme (VIL50302).

\bibliography{scibib}

\bibliographystyle{Science}

\end{document}